\documentclass[preprint]{aastex}

\slugcomment{Accepted by ApJ}

\shorttitle{The CME Structure in the EUV Passbands}

\shortauthors{Song et al.}

\begin{document}
%\title{On the Nature of the Dark Cavity of Solar Coronal Mass Ejections}
\title{The Structure of Solar Coronal Mass Ejections in the Extreme-Ultraviolet Passbands}
\author{H. Q. Song\altaffilmark{1}, J. Zhang\altaffilmark{2}, L. P. Li\altaffilmark{3}, Y. D. Liu\altaffilmark{4}, B. Zhu\altaffilmark{4}, B. Wang\altaffilmark{1}, R. S.
Zheng\altaffilmark{1}, and Y. Chen\altaffilmark{1}}

%\author{H. Q. Song\altaffilmark{1} et al.}

\affil{1 Shandong Provincial Key Laboratory of Optical Astronomy
and Solar-Terrestrial Environment, and Institute of Space
Sciences, Shandong University, Weihai, Shandong 264209, China}
\email{hqsong@sdu.edu.cn}

\affil{2 Department of Physics and Astronomy, George Mason
University, Fairfax, VA 22030, USA}

\affil{3 Key Laboratory of Solar Activity, National Astronomical
Observatories, Chinese Academy of Sciences, Beijing, 100101,
China}

%\affil{2 School of Atmospheric Sciences, Sun Yat-sen University,
%Zhuhai, Guangdong 519000, China}

\affil{4 State Key Laboratory of Space Weather, National Space
Science Center, Chinese Academy of Sciences, Beijing, 100190,
China}

%\affil{5 School of Astronomy and Space Science, Nanjing
%University, Nanjing, Jiangsu 210093, China}

%\affil{2 Department of Physics, Montana State University, Bozeman,
%MT 59717, USA}

%\affil{5 Yunnan Observatories, Chinese Academy of Sciences,
%Kunming, Yunnan 650216, China}

\begin{abstract}
So far most studies on the structure of coronal mass ejections
(CMEs) are conducted through white-light coronagraphs, which
demonstrate about one third of CMEs exhibit the typical three-part
structure in the high corona (\textit{e.g.}, beyond 2 $R_\odot$),
\textit{i.e.}, the bright front, the dark cavity and the bright
core. In this paper, we address the CME structure in the low
corona (\textit{e.g.}, below 1.3 $R_\odot$) through
extreme-ultraviolet (EUV) passbands and find that the three-part
CMEs in the white-light images can possess a similar three-part
appearance in the EUV images, \textit{i.e.}, a leading edge, a
low-density zone, and a filament or hot channel. The analyses
identify that the leading edge and the filament or hot channel in
the EUV passbands evolve into the front and the core later within
several solar radii in the white-light passbands, respectively.
What's more, we find that the CMEs without obvious cavity in the
white-light images can also exhibit the clear three-part
appearance in the EUV images, which means that the low-density
zone in the EUV images (observed as the cavity in white-light
images) can be compressed and/or transformed gradually by the
expansion of the bright core and/or the reconnection of magnetic
field surrounding the core during the CME propagation outward. Our
study suggests that more CMEs can possess the clear three-part
structure in their early eruption stage. The nature of the
low-density zone between the leading edge and the filament or hot
channel is discussed.

\end{abstract}


%\keywords{instabilities $-$ magnetic reconnection $-$ Sun:
%coronal mass ejections (CMEs) $-$ Sun: flares}

\section{Introduction}
Coronal mass ejections (CMEs) are one of the most energetic
explosions in the solar atmosphere, which can release large
quantities of magnetized plasmas, magnetic fluxes, and energetic
particles into the interplanetary space (Forbes et al. 2006; Chen
2011; Webb \& Howard 2012), and produce geomagnetic storms that
may adversely impact human high-technology systems around the
Earth (e.g., Gosling et al. 1991; Webb et al. 1994, 2000; Zhang et
al. 2003, 2007). Theoretically, CMEs result from the eruption of
magnetic flux ropes (MFRs), whenever the ropes are formed prior to
(Chen 1996; Lin \& Forbes 2000; Lin et al. 2004) or during
(Miki\'{c} \& Linker 1994; Antiochos et al. 1999) the eruptions.
MFR is a coherent magnetic structure with magnetic field lines
wrapping around its central axis, \textit{i.e.}, its field lines
exhibiting obvious twist (e.g., twist number $>$ 1). There is no
physical mechanism that can produce a CME from the corona without
involving an MFR. Observationally, CMEs are usually associated
with the eruption of filaments (\textit{i.e.}, prominences when
located near the solar limb, Webb \& Hundhausen 1987; Gopalswamy
et al. 2003) or hot channels (Nindos et al. 2015; Zhang et al.
2015). Filaments are dense and cold materials trapped at the MFR
dips, \textit{i.e.}, sites where the field lines are locally
horizontal and curved upward (Kippenhahn \& Schl\"uer, 1957). Hot
channels refer to the high-temperature coronal structure revealed
in the 131 or 94 \AA\ passbands (Zhang et al. 2012), and they can
appear as the hot blobs if observed along their axis due to the
projection effect (Cheng et al. 2011; Song et al. 2014a, 2014b).
Many researches support that the hot channels act as the proxy of
MFRs (Zhang et al. 2012; Patsourakos et al. 2013; Cheng et al.
2014; Song et al. 2015). Therefore, the eruption of MFRs in
theories usually manifests as the ejection of filaments or hot
channels in observations.

Understanding various aspects of CMEs is of important
significance, among which one basic issue is their structure that
provides a crucial clue to investigate the eruption process. A
recent statistical study classified CMEs into five categories
based on their appearance characteristics in the white-light
coronagraphic images (Vourlidas et al. 2013), including the
three-part structure (CMEs with a bright front, a dark cavity, and
a bright core), loop (CMEs with a bright loop but lacking a cavity
and/or a core), jet (narrow CMEs with angular width less than 40
degrees), outflow (CMEs wider than jets, without clear loop front
or cavity), and failed (CMEs that disappear in the outer corona).
The three-part structure has been considered as the archetypical
configuration in the research field, though previous studies
revealed that only about one third of CMEs possess this appearance
(e.g., Illing \& Hundhausen 1985). Song et al. (2017) demonstrated
that at least two factors can reduce the probability of CMEs
exhibiting the three-part structure, including the observational
time or altitude (evolution effect) and perspective (projection
effect). They claim that more three-part CMEs could be observed in
the early stage of CME eruptions (viz in the low corona) and/or
with suitable observational viewpoint. To further examine the
validity of these claims is our first motivation to investigate
the CME structure in the low corona through EUV passbands.

As the brightness in white-light coronagraphs is proportional to
the electron column density (e.g., Hayes et al. 2001), the
three-part structure corresponds to a high-low-high density
sequence. When the MFR, with dense filament located in its
trailing part, lifts off from the source region, it will expand
and compress its overlying loops successively (e.g., Chen 2009),
then the background plasmas can pile up along the MFR front border
and evolve into the CME bright front (Forbes, 2000). Therefore,
the three-part structure of CMEs has been explained as the
manifestations of coronal plasma pileup (high density), MFR (low
density), and filament (high density) sequentially for several
decades. Recently, the interpretation of the CME core is
challenged (Howard et al. 2017; Song et al. 2017). Howard et al.
(2017) conducted a survey based on 42 CMEs all with the three-part
structure, which illustrated that $\sim$69\% of the events are not
associated with any eruptive filament. They speculated that the
CME core is produced by the geometric projection of a twisted MFR.
Song et al. (2017) clearly demonstrated that the hot channel MFR
corresponds to the bright core through a filament-unrelated CME
from both edge-on and face-on perspectives. Therefore, both the
filament and MFR (e.g., hot channel) can be observed as the CME
core in the coronagraphs. Song et al. (2019) further presented a
three-part CME with both a sharp and a fuzzy core, which
correspond to the filament and the MFR, respectively. Since the
MFR may appear as the CME core in the new scenario, one key
question arises naturally, \textit{i.e.}, what is the nature of
the CME cavity in the white-light images? It might correspond to
different part of the MFR due to projection effect (e.g., Howard
et al. 2017). Alternatively, the cavity might be just a
low-density zone without helical fields (Song et al. 2017, 2019).
To shed more light on this issue is our second motivation to
investigate the CME structure in the EUV passbands.

The white-light coronagraphs capture the Thomson-scattered light
from the free electrons in the corona, which is dependent on the
electron density as mentioned, while the EUV passbands are
dependent on both the density and temperature of the coronal
plasma (e.g., Del Zanna \& Mason 2018). In the meantime, the
white-light intensities are more sensitive to coronal features
near the plane of sky (POS, e.g., Vourlidas \& Howard 2006), while
the EUV passbands are less preferentially sensitive to features
based on their locations relative to the POS. Therefore, the
density structures in the white-light and the EUV passbands can be
compared and correlated each other more straightforwardly for limb
CMEs. We select four CMEs originating near the limb in this paper,
including two (two) events with (without) the obvious cavity in
the coronagraphs, which are associated with the eruption of a
filament and a hot channel, respectively. The instruments are
introduced in Section 2, and the observations and results are
presented in Section 3. Section 4 is our summary, which is
followed by discussions in Section 5.

\section{Instruments}
The EUV data sets are provided by three instruments, including the
Atmospheric Imaging Assembly (AIA; Lemen et al. 2012) on board the
\textit{Solar Dynamics Observatory (SDO)}, the Extreme Ultraviolet
Imager (EUVI; Howard et al. 2008) on board the \textit{Solar
Terrestrial Relations Observatory (STEREO)}, as well as the Solar
Ultraviolet Imager (SUVI; Seaton \& Darnel 2018) on board the
\textit{GOES-16}. The white-light coronagraphs COR1 (Howard et al.
2008) on board the \textit{STEREO}, and the Large Angle and
Spectrometric Coronagraph (LASCO; Brueckner et al. 1995) on board
the \textit{Solar and Heliospheric Observatory (SOHO)} are used to
analyze the CME structure in the high corona. The \textit{SOHO},
\textit{SDO}, and \textit{GOES-16} observe the Sun from the Earth
viewpoint. The \textit{STEREO} consists of twin spacecraft
orbiting the Sun, one ahead of (A) and the other behind (B) the
Earth, which can provide observations from different perspectives
as each one separates from the Earth by $\sim$22${^\circ}$ every
year in heliocentric longitude.

The AIA records the low corona with the FOV (field of view) of 1.3
$R_\odot$, cadence of 12 s and resolution of 0.\arcsec6 per pixel
in seven EUV passbands. The SUVI images the solar corona with an
FOV larger than 1.6 $R_\odot$ in the horizontal direction in six
EUV passbands. The EUVI provides the solar EUV images at four
wavelength with an FOV of 1.7 $R_\odot$, partially overlapping
with that of COR1 (1.4--4 $R_\odot$), which enables us to observe
the object continuously from the solar surface to the outer
corona. The LASCO possesses an FOV covering 2.2--30 $R_\odot$ (C2:
2.2--6 $R_\odot$; C3: 4--30 $R_\odot$).

\section{Observations and Results}
\subsection{Two CMEs with Obvious Cavity}
In this subsection, two CMEs with typical three-part structure in
the white-light images are analyzed. One occurred on 2013
September 24, associated with a filament eruption, and the other
erupted on 2015 February 9, associated with a hot channel
eruption.

\subsubsection{The 2013 September 24 Event}
A CME was recorded by the LASCO on 2013 September 24 as shown in
Figure 1(a1), which exhibits a clear and typical three-part
structure. The CME first appeared in the C2 FOV at 20:36 UT and
propagated outward at a linear speed of 919 km s$^{-1}$ (CDAW:
$https://cdaw.gsfc.nasa.gov/$) near the Sun. Through inspecting
the SDO/AIA observations, we can correlate this CME with a
filament eruption unambiguously. Figure 1(a2) presents one
snapshot of the filament eruption provided by the AIA
running-difference image at 193 \AA\ wavelength, which
demonstrates that the event already possesses the three-part
appearance in the early stage, \textit{i.e.}, the leading edge as
marked with blue dots, the filament as denoted with the red arrow,
and a zone between them can be distinguished. Please see the
accompanying animation to examine the eruption process in the EUV
passband.

The \textit{STEREO-B} is $\sim$139${^\circ}$ east of the Earth
during this eruption. Both the EUVI B and COR1 B record the CME
and its source region. Figure 2(a) presents the EUVI B 195 \AA\
running-difference image at 20:16 UT, which also displays the
three-part appearance with the blue and red dots depicting the
leading edge and filament, respectively. These dots are replotted
in the COR1 image as shown in Figure 2(b). It is straightforward
that the leading edge in the EUV passband corresponds to the CME
front in the white-light coronagraph. At 20:16 UT, the filament as
denoted with the red dots is still blocked by the occulter as
shown in Figure 2(b), hence no CME core is observed by the COR1 B
at this time. Twenty minutes later, the filament ascends to a
higher altitude as presented in Figure 2(c). The filament front is
depicted with red dots again and the dots are superimposed on the
COR1 image recorded at the same time (Figure 2(d)), which clearly
demonstrates that the filament corresponds to the CME core. A
similar situation was obtained by overlapping EUVI and COR1 (Liu
et al. 2010; Liu et al. 2014). Therefore, we conclude that the
leading edge and filament observed in the EUV passband evolve into
the bright front and core of three-part CME in the white-light
images. Naturally, the zone between the leading edge and the
filament should evolve into the dark cavity of the CME in the
white-light images, illustrating the zone with lower density. Note
the cyan squares in the right panels depict the FOV of left
panels.

\subsubsection{The 2015 February 9 Event}
Another CME with the typical three-part structure occurred on 2015
February 9 and was recorded by the LASCO as displayed in Figure
1(b1). The CME first appeared in the C2 FOV at 23:24 UT and moved
outward at the linear speed of 1106 km s$^{-1}$ (CDAW) in the
LASCO FOV. After inspecting the observations of LASCO and AIA, we
can correlate this CME with a hot channel eruption. Figure 1(b2)
is a composite observation of AIA 193 \AA\ running-difference
image (gray) and 131 \AA\ direct image (cyan). The leading edge
(193 \AA) and the hot channel (131 \AA) are depicted with blue and
red dots, respectively, which shows that a zone exists between the
leading edge and the hot channel. Please check the accompanying
animation to observe the three-part configuration in the EUV
passbands clearly. As the hot channel density is higher than its
surrounding environment as revealed by the differential emission
measure analysis (Cheng et al. 2012), \textit{i.e.}, the zone
density is lower compared to both the hot channel and the piled-up
leading edge. Therefore, for this CME induced by the hot channel
eruption, we also believe that the three-part configuration in the
EUV images evolves into the three-part structure in the
white-light images correspondingly, which means the hot channel
evolves into the bright core of the CME as suggested in Song et
al. (2017).

\subsection{Two CMEs without Obvious Cavity}
In this subsection, we analyze two CMEs without obvious cavity
existing between the front and core in the white-light images,
which are also associated with the eruptions of a filament and a
hot channel, respectively. One occurred on 2014 January 6, and the
other on 2017 September 10.

\subsubsection{The 2014 January 06 Event}
The LASCO recorded a CME on 2014 January 6 that first appeared in
the C2 FOV at 08:00 UT and moved in the coronagraphic FOV at a
linear speed of 1402 km s$^{-1}$ (CDAW). The CME does not exhibit
the clear three-part structure in the white-light image as no
obvious dark cavity can be distinguished as displayed in Figure
3(a1). One interesting phenomenon is that the CME front exhibits a
local deformation, \textit{i.e.,} a bulge appears at the front as
denoted with the yellow arrow, which can be observed more clearly
in the running-difference image (Figure 3(a2)).

The \textit{STEREO-B} is $\sim$153${^\circ}$ east of the Earth
during this eruption. Through inspecting both the white-light and
EUV observations, the source region can be identified
unambiguously. The CME results from a filament eruption that is
recorded by the AIA and EUVI B simultaneously from two viewpoints.
Figure 3(a3) presents the AIA 193 \AA\ running-difference image,
demonstrating the three-part appearance, \textit{i.e.,} the
leading edge, the filament as denoted with the red arrow, and the
zone between them. The EUVI B 195 \AA\ observation presents the
similar result as shown in Figure 3(a4), where the filament is
depicted with the red arrow. The leading edge, the filament and
the zone between them in the EUV images (panel (a3)) should evolve
into the bright front, bright core and dark cavity in the
white-light images, respectively. However, no obvious dark cavity
is recorded in the coronagraph as shown in panel (a1). This will
be discussed later with the next event together.

\subsubsection{The 2017 September 10 Event}
This event was recorded by the \textit{SOHO} and \textit{STEREO-A}
from two perspectives on 2017 September 10 when the
\textit{STEREO-A} was 128${^\circ}$ east of the Earth, which was
accompanied by an X8.2 class flare and has been used to
investigate various aspects of CMEs (Yan et al. 2018; Veronig et
al. 2018 and references therein). The CME first appeared in the C2
FOV at 16:00 UT at a high linear speed of 3163 km s$^{-1}$ (CDAW).
The LASCO observation shows that the CME does not possess the
obvious dark cavity either as displayed in Figure 3(b1). The CME
front also exhibits a bulge as depicted with the yellow arrow,
which is more distinguishable in the running-difference image
(Figure 3(b2)).

Through inspecting the LASCO and AIA observations, we can find the
source region without ambiguity. Figure 3(b3) presents the AIA 131
\AA\ observations, clearly showing that the CME is driven by the
eruption of a hot blob. Due to the AIA limited FOV, partial blob
and the leading edge are not imaged by the AIA at 15:55:30 UT. We
acquire the blob outline based on the observation at 15:55:06 UT
as marked with red dots in the inset panel, assuming that the blob
does not change its morphology significantly. At the same time,
the leading edge can be observed clearly in the SUVI FOV through
the 195 \AA\ passband as shown in Figure 3(b4), where the hot blob
outline recorded by the AIA is replotted with red dots. This
figure clearly demonstrates that the CME also possesses the
three-part appearance in the EUV passbands. The bulge is formed in
the propagation direction of the expanding hot blob, which moves
much faster than the CME front, and thus protrudes from the bottom
of the CME front (Veronig et al. 2018). This illustrates the
low-density zone in the low corona can be compressed significantly
during propagation outward.

Figure 4(a) displays the EUVI A 195 \AA\ observation at 15:58 UT,
only showing the leading edge as depicted with the blue dots.
Since the hot blob temperature is very high (beyond 10 MK, see
Cheng et al. 2018), it is unobservable in the 195 \AA\ passband.
Please check the accompanying animation to examine the eruption in
the EUV passband. However, the high density characteristics of
both the leading edge (blue dots) and the hot blob (red dots) can
be revealed by the COR1 observation simultaneously as shown in
Figure 4(b) after they entered into the COR1 FOV at 16:05 UT. Note
the central part of the bright front is not clear enough to be
identified in the static image, please check the animation for the
continuous observations in the white-light passband. This
demonstrates again that the leading edge and the hot blob evolve
into the bright front and the core of CME, respectively.

Figure 5 presents six running-difference images recorded by the
LASCO/C3. The bulge at the CME front keeps obvious as denoted with
the yellow arrow in each panel, and locates in the propagation
direction of the expanding core as depicted with the red arrows.
In the meantime, as the CME expansion speed decreases over time
(Liu et al. 2019), the separation between the driven shock and the
CME front appeared gradually at both flanks of the CME as depicted
with the blue arrows. This phenomenon has been reported in the EUV
passband (Cheng et al. 2012), demonstrating that the EUV leading
edge can contain two components, i.e., the pile-up plasma
(corresponding to the bright front of CME in the coronagraphs) and
the EUV wave (or shock when the CME speed is high enough).

\section{Summary}
In this paper, we selected four limb events to study the CME
structure at their initial eruption stage through the EUV
passbands. The observations demonstrated that all the four CMEs
can possess the three-part appearance in the EUV images, wherever
they were associated with the eruption of a filament or a hot
channel MFR, and whether they had obvious cavity or not in the
white-light images. If the three-part appearance in the low corona
observed through the EUV passbands can be imaged by the
coronagraphs at the same time, the white-light data will also
record a three-part CME with the obvious cavity. Our study further
confirmed that both the filament and the hot channel can appear as
the bright core of CMEs in the coronagraphs. Our observations
illustrated that the low-density zone (dark cavity) can decrease
gradually in size (Subsection 3.2) when the CME propagates outward
due to the core expansion and/or reconnection, which obscures or
eliminates the three-part structure in the high corona. Therefore,
we suggest that more CMEs (not only about one third as usually
cited) possess the typical three-part structure in the low corona
(e.g., below 1.3 $R_\odot$). This is consistent with the cavity
surveys in both EUV and white-light passbands, which demonstrated
that nearly 80\% of the surveyed days had one or more EUV cavity
(Forland et al. 2013), versus 10\% of the white-light days
surveyed (Gibson et al. 2006). More comprehensive statistics
should be conducted to get the exact percentage of the three-part
CMEs in the low corona.

\section{Discussions on the Nature of the Low-density Zone (Dark Cavity)}
Based on the traditional explanation of the three-part structure
of CMEs, \textit{i.e.} the bright core and the dark cavity
correspond to the filament and the MFR, respectively, the MFR will
expand and the cavity volume should increase correspondingly
during the propagation, instead of decreasing as observed in
Subsection 3.2. On the other hand, based on the new scenario of
the CME three-part structure, in which the bright core and the
dark cavity correspond to the MFR and the low-density zone,
respectively, it is straightforward to understand the decrease of
the dark cavity volume. Both the expansion of the bright core
(MFR) and the reconnection of magnetic field surrounding the core
can lead to the decrease of the dark cavity in size during the CME
propagation outward. Therefore, the zone between the leading edge
and the hot channel or filament observed in the EUV passband
(\textit{i.e.}, in the early eruption stage or in the low corona)
might not totally correspond to the MFR structure.

The existence of a low-density zone between the overlying loops
and the hot channel MFR has been displayed clearly through a
failed eruption (see Figure 1 of Song et al. 2014b), which
demonstrates that the MFR does not occupy all the space below the
leading edge in the eruption stage. The low-density zone might be
occupied by sheared field lines, different from the overlying
loops that are closer to a potential field. During the eruption,
the MFR will expand and ascend, and the field lines in the cavity
could be transformed into the shell part of the MFR through
reconnection occurring in the current sheet beneath the CME (Lin
\& Forbes 2000; Lin et al. 2004; Zuccarello et al. 2012). The
reconnection also injects plasmas into the MFR along the
reconnected field lines, and the newly formed shell part of the
MFR might possess relatively low or high density depending on the
injections and appear as part of the dark cavity or bright core
correspondingly. Hence the so-called low-density zone (cavity)
could keep existing (Figure 1) or disappear gradually (Figure 3)
during the evolution process.

After the CMEs propagate into the high corona (e.g., beyond 2
$R_\odot$), their cavity might correspond to the MFR totally. We
propose one preliminary method to judge the nature of the cavity
based on its evolution trend. If the cavity is the MFR, the cavity
will continue to expand during the propagation outward and exist
for a long time. Eventually it can be detected through the in situ
measurements that prove the cavity to be the MFR (Liu et al. 2010;
Howard \& DeForest 2012). On the other hand, if the cavity is
still a low-density zone, its volume might continue to decrease
gradually due to the MFR expansion and/or magnetic reconnection
(Song et al. 2017; Song et al. 2019). The results in this
qualitative study provide one explanation why some CMEs do not
exhibit the clear three-part structure (\textit{i.e.}, no obvious
cavity) in the high corona. More quantitative analyses and
numerical simulations are necessary in the future.

\acknowledgments We thank the anonymous referee for the comments
and suggestions that helped to improve the original manuscript. We
are grateful to Drs. Xin Cheng (NJU), Peng-Fei Chen (NJU), Jun Lin
(YNO), Gang Li (UAH), and Bo Li (SDU) for their valuable
discussions. We acknowledge the use of data from the \textit{SDO},
\textit {SOHO}, \textit{STEREO}, and \textit{GOES} missions. This
work is supported by the Shandong Provincial Natural Science
Foundation (JQ201710), the NSFC grants U1731102, U1731101,
41331068, 11790303, and 11790300, and the CAS grants XDA-17040505
and XDA-15010900.

\clearpage

\begin{figure}
\epsscale{0.9} \plotone{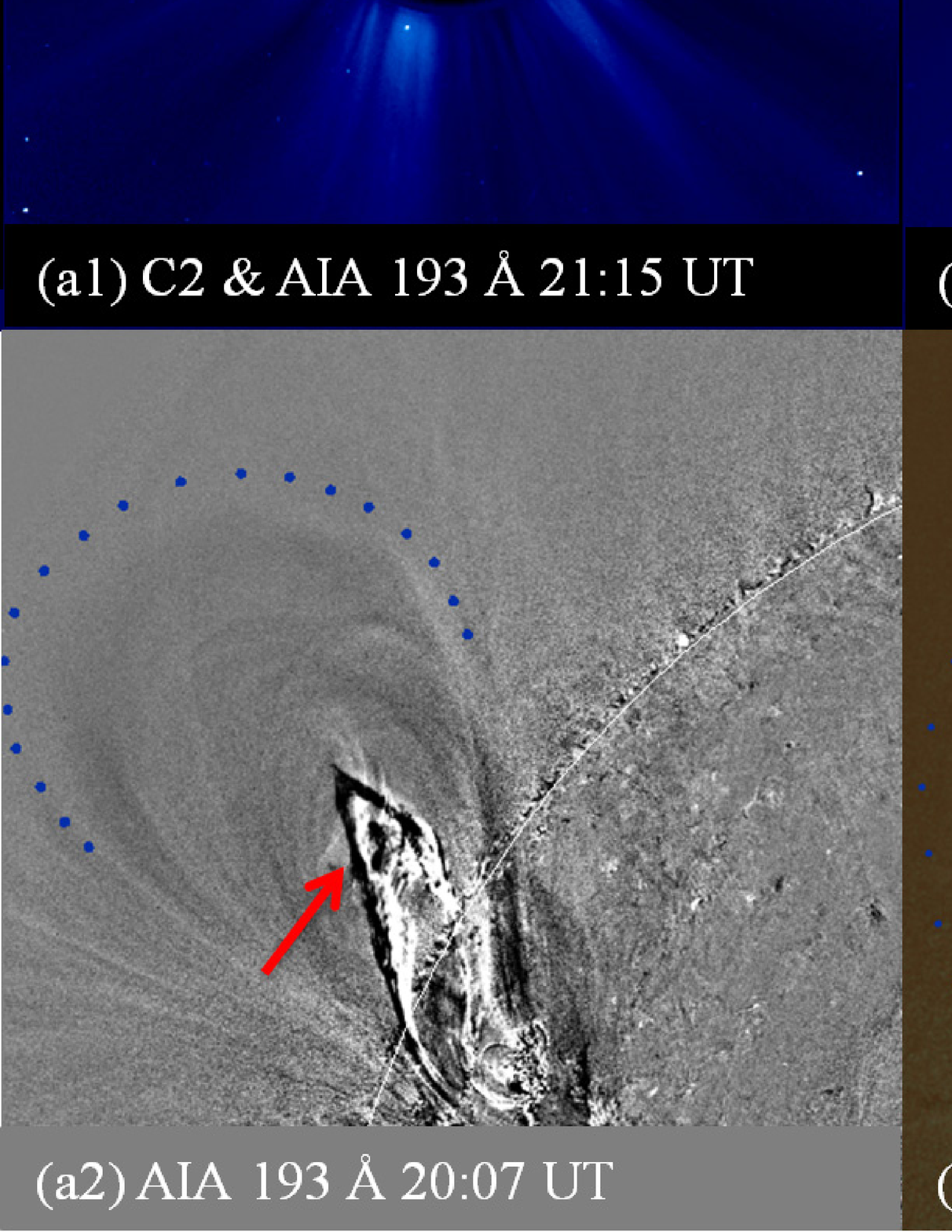} \caption{Two CMEs with obvious
cavity in the coronagraph. Left: The event on 2013 September 24.
(a1) The LASCO/C2 direct image with the AIA 193 \AA\ image
superimposed. (a2) The AIA 193 \AA\ running-difference image,
showing the leading edge (blue dots) and the filament as denoted
with the red arrow. Right: The event on 2015 February 9. (b1) The
LASCO/C2 direct image with the AIA 193 \AA\ image superimposed.
(b2) The composite image of AIA 193 running-difference image
(gray) and 131 \AA\ direct image (cyan), showing the leading edge
(blue dots) and the hot channel as denoted with the red dots.
Panel (a2) is accompanied by an animation that displays the
filament eruption process from 19:40 UT to 20:10 UT with a
duration of 2 seconds. Panel (b2) is accompanied by the same
animation that also displays the hot channel eruption process from
22:50 UT to 23:14 UT. The animation shows the Panel (a2)/(b2) at
the left/right side on each frame of the video. \label{Figure 1}}
\end{figure}

\begin{figure}
\epsscale{1.0} \plotone{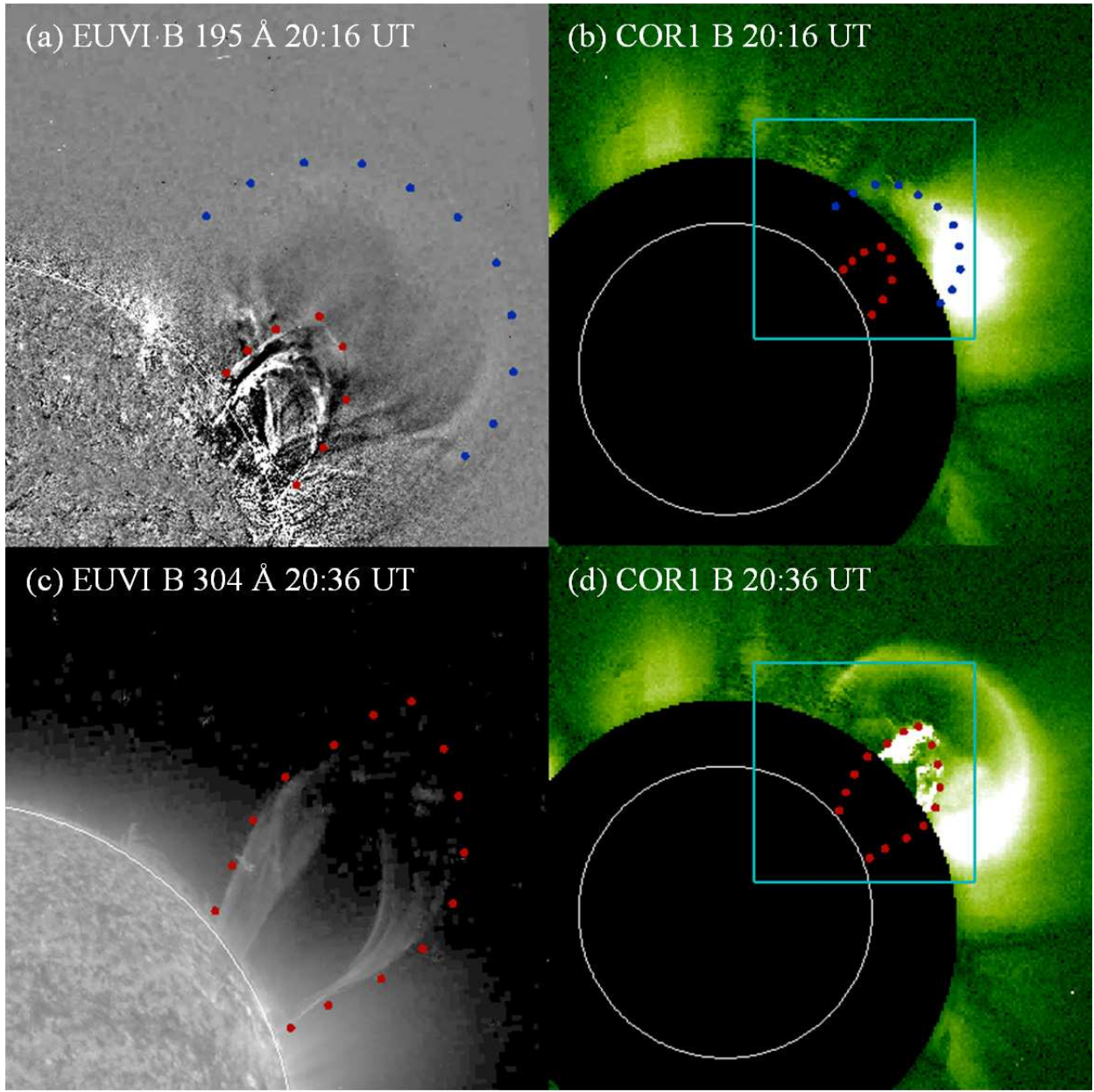} \caption{Observations of the
event on 2013 September 24 by the EUVI B and COR1 B. (a) The
running-difference image of EUVI B 195 \AA. (b) and (d) The direct
images of COR1 B, showing the CME morphology. (c) The direct image
of 304 \AA. Panel (a) is accompanied by an animation that displays
the complete eruption process recorded by the EUVI B. The
animation starts at 19:41 UT and ends at 20:26 UT with a duration
of 2 seconds. \label{Figure 2}}
\end{figure}

\begin{figure}
\epsscale{0.6} \plotone{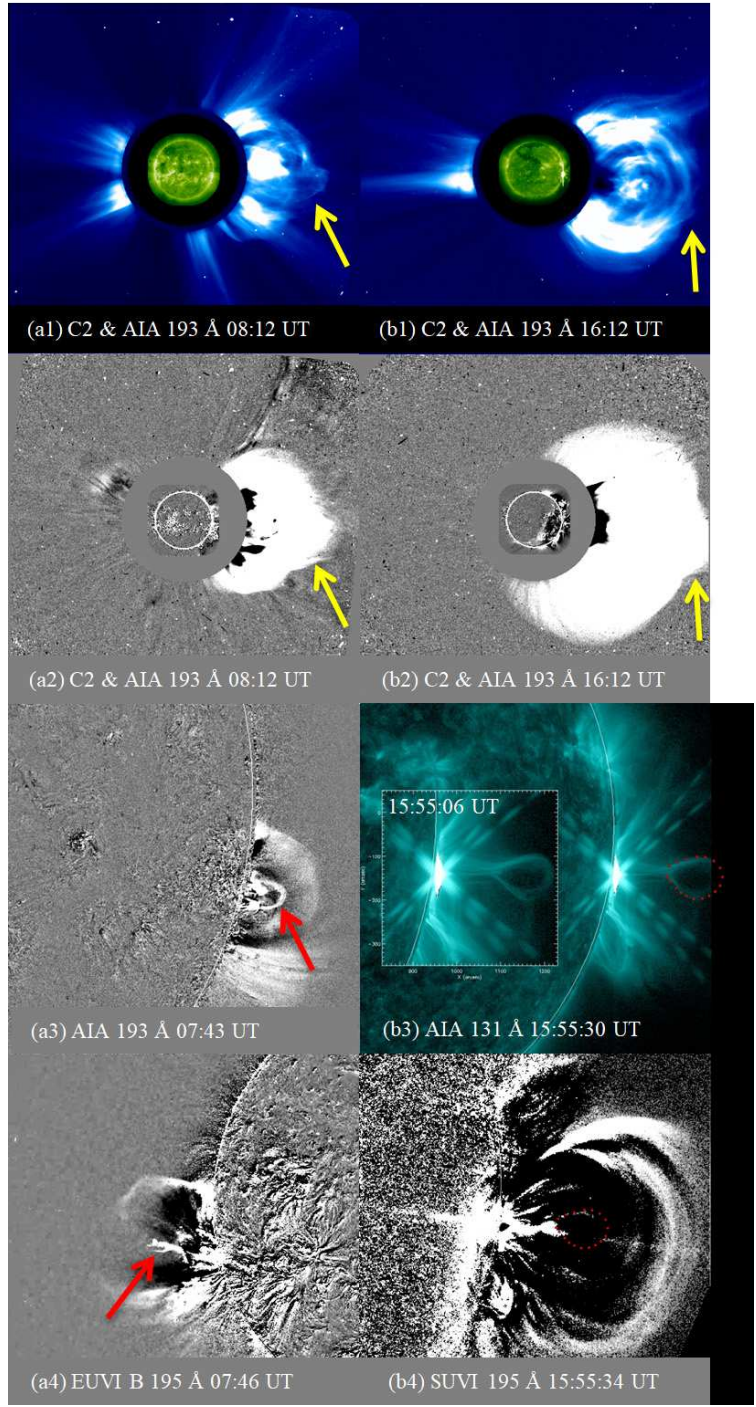} \caption{Two CMEs wihtout
obvious cavity in the coronagraph. Left: The event on 2014 January
6. (a1) The composite image of the LASCO/C2 and AIA 193 \AA.
(a2)-(a4) The running-difference image of the LASCO/C2, AIA 193
\AA, and EUIV B 195 \AA, respectively. Right: The event on 2017
September 10. (b1) The composite image of the LASCO/C2 and AIA 193
\AA. (b2) The running-difference image of the LASCO/C2. (b3) The
direct image of AIA 131 \AA. (b4) The running-difference image of
the SUVI 195 \AA\ with the 131 \AA\ hot blob outline overplotted.
\label{Figure 3}}
\end{figure}

\begin{figure}
\epsscale{0.7}\plotone{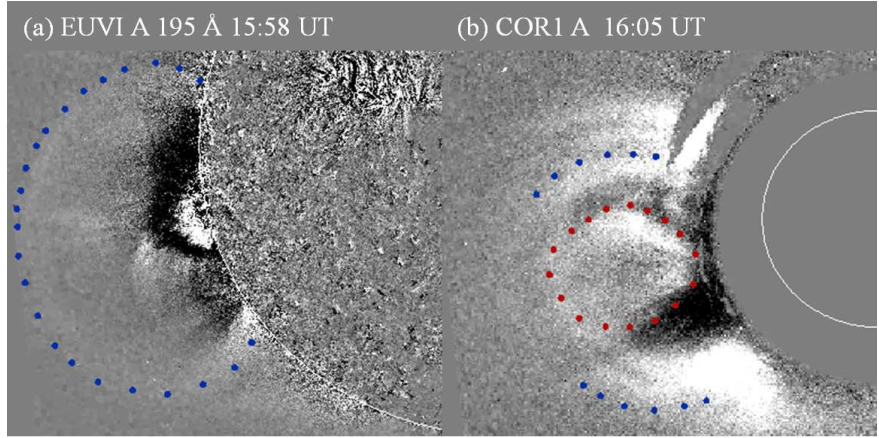} \caption{The \textit{STEREO-A}
observation of the event on 2017 September 10. (a) The
running-difference image of EUVI A 195 \AA. The blue dots denote
the leading front. (b) The running-difference image of COR1 A. The
blue and red dots depict the leading edge and the core,
respectively. Panel (a) is accompanied by an animation that
displays the eruption process from 15:48 UT to 16:10 UT with a
duration of 2 seconds. Panel (b) is accompanied by the same
animation that displays the eruption process from 15:50 UT to
16:10 UT. The animation shows the Panel (a)/(b) at the left/right
side on each frame of the video. \label{Figure 4}}
\end{figure}

\begin{figure}
\epsscale{0.8}\plotone{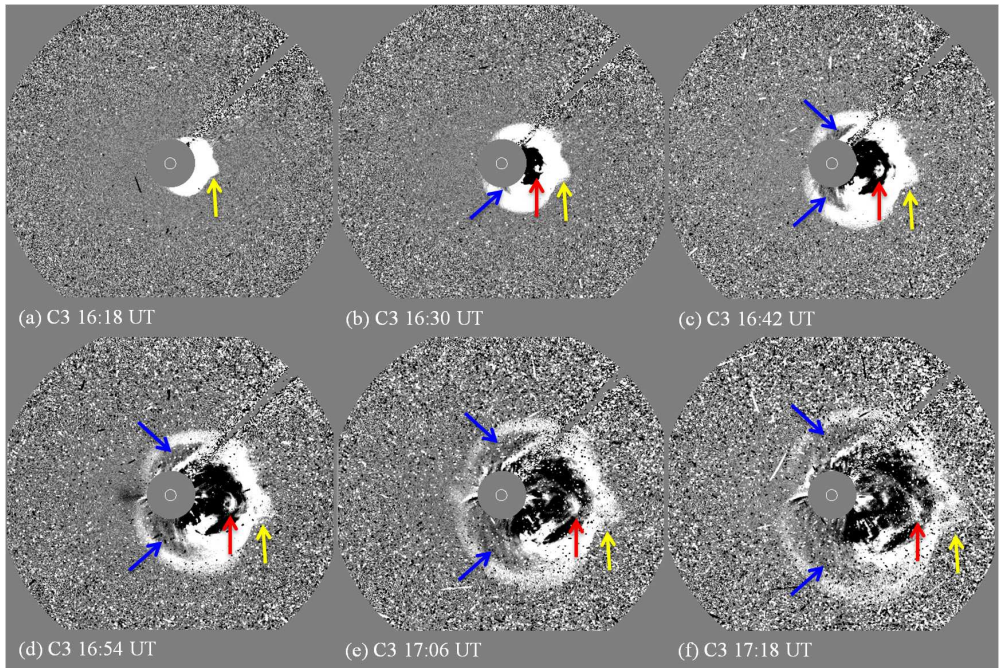} \caption{The running-difference
images of LASCO/C3 on the 20170910 event. The bulge, core, and
separation between the shock and the CME front are depicted with
the yellow, red, and blue arrows, respectively. The white circle
in each panel delineates the solar limb. \label{Figure 5}}
\end{figure}

%\begin{figure}
%\epsscale{0.8}\plotone{Fig6.eps} \caption{Schematic drawings of
%the traditional (a1) and the new (b1) scenarios explaining the
%three-part structure of CMEs. The former needs the filament moving
%inside the MFR to interpret the bulge of CME front (a2), which
%seems impossible. The later can explain the bulge
%straightforwardly (b2). The red solid line depicts the MFR
%boundary in each panel. \label{Figure 6}}
%\end{figure}


\begin{thebibliography}{}
%\bibitem[Alexander et al. (2006)]{Alexander 2006}
%Alexander, D., Liu, R., \& Gilbert, H. R. 2006, \apj, 653, 719

%\bibitem[Amari et al.(2014)]{2014Natur.514..465A}
%Amari, T., Canou, A.,\& Aly, J.-J.\ 2014, \nat, 514, 465

%\bibitem[Antiochos et al. (1999)]{Antiochos 1999}
%Antiochos, S. K., Devore, C. R., \& Klimchuk, J. A. 1999, \apj,
%510, 485

%\bibitem[Aschwanden \& Boerner (2011)]{Aschwanden 2011}
%Aschwanden, M. J., \& Boerner, P. 2011, \apj, 732, 81

%\bibitem[Aulanier et al. (2012)]{Aulanier 2012}
%Aulanier, G., Janvier, M., \& Schmieder, B. 2012, \aap, 543, A110

%\bibitem[Badnell et al. (2003)]{Badnell 2003}
%Badnell, N. R., O'Mullane, M. G., Summers, H. P., et al. 2003, \aap, 406, 1151

%\bibitem[Benz (2008)]{Benz 2008}
%Benz, A. O. 2008, Living Rev. Solar Phys., 5, 1

\bibitem[Brueckner et al.(1995)]{1995SoPh..162..357B}
Brueckner, G.~E., Howard, R.~A., Koomen, M.~J., et al.\ 1995,
\solphys, 162, 357

%\bibitem[Burlaga et al. (1981)]{Burlaga 1981}
%Burlaga, L., Sittler, E., Mariani, F., \& Schwenn, R. 1981, \jgr,
%86, 6673

%\bibitem[Byrne et al.(2014)]{2014SoPh..289.4545B}
%Byrne, J.~P., Morgan, H., Seaton, D.~B., Bain, H.~M., \& Habbal,
%S.~R.\ 2014, \solphys, 289, 4545

%\bibitem[Capannolo et al. (2017)]{Capannolo 2017}
%Capannolo, L., Opher, M., Kay, C., \& Landi, E. 2017, \apj, 839,
%37

%\bibitem[Carmichael(1964)]{1964NASSP..50..451C}
%Carmichael, H.\ 1964, NASA Special Publication, 50, 451


%\bibitem[Chen et al.(2014)]{2014ApJ...794..149C}
%Chen, B., Bastian, T.~S.,\& Gary, D.~E.\ 2014, \apj, 794, 149

\bibitem[Chen (1996)]{Chen 1996}
Chen, J. 1996, \jgr, 101, 27499

%\bibitem[Chen (2000)]{Chen 2000}
%Chen, J., Santoro, R. A., Krall, J., et al. 2000, \apj, 533, 481

\bibitem[Chen (2009)]{CME front formation}
Chen, P. F.\ 2009, \apjl, 698, L112

\bibitem[Chen(2011)]{2011LRSP....8....1C}
Chen, P.~F.\ 2011, LRSP, 8, 1


%\bibitem[Chen et al.(2007a)]{2007ApJ...665.1421C}
%Chen, Y., Hu, Y.~Q., \& Sun, S.~J.\ 2007a, \apj, 665, 1421


%\bibitem[Chen et al.(2007b)]{2007AdSpR..40.1780C}
%Chen, Y., Hu, Y.~Q., \& Xia, L.~D.\ 2007b, Adv. Space Res., 40,
%1780

%\bibitem[Chen et al.(2006)]{Chen 2006}
%Chen, Y., Li, G. Q., \& Hu, Y.~Q.\ 2006, \apj, 649, 1093

\bibitem[Cheng et al. (2014a)]{Cheng 2014 hc being the MFR}
Cheng, X., Ding, M. D., Guo, Y., et al. 2014, \apj, 780, 28

%\bibitem[Cheng et al.(2014b)]{2014ApJ...789L..35C}
%Cheng, X., Ding, M.~D., Zhang, J., et al.\ 2014, \apjl, 789, L35

%\bibitem[Cheng et al.(2014c)]{2014ApJ...789L..93}
%Cheng, X., Ding, M.~D., Zhang, J., et al.\ 2014, \apjl, 789, 93

\bibitem[Cheng et al. (2018)]{xin cheng}%{a20170910 event a}
Cheng, X., Li, Y., Wan, L. F., et al. 2018, \apj, 866, 64

%\bibitem[Cheng et al.(2013)]{2013ApJ...763..43}
%Cheng, X., Zhang, J., Ding, M.~D., Liu, Y., \& Poomvises, W. 2013,
%\apj, 763, 43

%\bibitem[Cheng et al. (2010)]{Cheng 2010}
%Cheng, X., Zhang, J., Ding, M. D., \& Poomvises, W. 2010, \apj, 712, 752

%\bibitem[Cheng et al.(2013)]{2013ApJ...769L..25C}
%Cheng, X., Zhang, J., Ding, M.~D., et al.\ 2013, \apjl, 769, L25

\bibitem[Cheng et al. (2011)]{Cheng 2011 Hot blob formation during eruption}
Cheng, X., Zhang, J., Liu, Y., \& Ding, M. D. 2011, \apj, 732, L25

\bibitem[Cheng et al. (2012)]{Cheng 2012}
Cheng, X., Zhang, J, Saar, S. H., \& Ding, M. D. 2012, \apj, 761,
62

%\bibitem[Cohen et al. (2010)]{Cohen et al. 2010}
%Cohen, O., Attrill, G. D. R., Schwadron, N. A., et al. 2010, \jgr,
%115, A10104

%\bibitem[Cremades et al. (2006)]{Cremades 2006}
%Cremades, H., Bothmer, V., \& Tripathi, D.\ 2006, Adv. Space Res.,
%38, 461

%\bibitem[D{\'e}moulin (2008)]{Demoulin 2008}
%D{\'e}moulin, P.\ 2008, Annales Geophysicae, 26, 3113

%\bibitem[D{\'e}moulin\& Aulanier(2010)]{2010ApJ...718.1388D}
%D{\'e}moulin, P., \&Aulanier, G.\ 2010, \apj, 718, 1388

\bibitem[Del Zanna \& Mason (2018)]{EUV emission on tem and density}
Del Zanna, G., \& Mason, H. E. 2018, LRSP, 15, 5

%\bibitem[Del Zanna et al. (2011)]{Del Zanna 2011}
%Del Zanna, G., O'Dwyer, B., \& Mason, H. E. 2011, \aap, 535, A46

%\bibitem[Doschek et al. (1980)]{Doschek 1980}
%Doschek, G. A., Feldman, U., Kreplin, R. W., \& Cohen, L. 1980, \apj, 239, 725

%\bibitem[Fan (2005)]{Fan 2005}
%Fan, Y. H. 2005, \apj, 630, 543

%\bibitem[Fan (2016)]{Fan 2016}
%Fan, Y. H. 2016, \apj, 824, 93

%\bibitem[Fan\& Gibson(2007)]{2007ApJ...668.1232F}
%Fan, Y., \& Gibson, S.~E.\ 2007, \apj, 668, 1232

%\bibitem[Feldman et al. (1980)]{Feldman 1980}
%Feldman, U., Doschek, G. A., Kreplin, R. W., \& Mariska, J. T. 1980, \apj, 241, 1175

\bibitem[Forbes (2000)]{Forbes 2000}
Forbes, T.~G., \ 2000, \jgr, 105, 23153

%\bibitem[Forbes\& Isenberg(1991)]{1991ApJ...373..294F}
%Forbes, T.~G., \& Isenberg, P.~A.\ 1991, \apj, 373, 294

\bibitem[Forbes et al. (2006)]{SSR CME review}
Forbes, T.~G., Linker, J. A., Chen, J., et al. 2006, \ssr, 123,
251

%\bibitem[Forbes \& Priest(1984)]{Forbes 1984}
%Forbes, T. G., \& Priest, E. R. 1984, in Solar Terrestrial
%Physics: Present and Future, ed. D. M. Butler \& K. Papadopoulus
%(Washington: NASA), 1

%\bibitem[Forbes\& Priest(1995)]{1995ApJ...446..377F}
%Forbes, T.~G., \& Priest, E.~R.\ 1995, \apj, 446, 377

%\bibitem[Forbes (2000)]{Forbes 2000}
%Forbes, T. G. 2000, \jgr, 105, 23153

%\bibitem[Forbes \& Isenberg (1991)]{Forbes 1991}
%Forbes, T. G., \& Isenberg, P. A. 1991, \apj, 373, 294

\bibitem[Forland et al. (2013)]{Forland et al. 2013 coronal cavity survey}
Forland, B. C., Gibson, S. E., Dove, J. B., Rachmeler, L. A., \&
Fan, Y. 2013, \solphys, 288, 603

%\bibitem[Gibson \& Fan (2006)]{Gibson 2006a}
%Gibson, S. E., \& Fan, Y. 2006, \apj, 637, L65

\bibitem[Gibson et al. (2006)]{Gibson 2006b}
Gibson, S. E., Foster, D., Burkepile, J., de Toma, G., \& Stanger,
A. 2006, \apj, 641, 590

%\bibitem[Gilbert et al. (2001)]{Gilbert 2001}
%Gilbert, H. R., Holzer, T. E., Low, B. C., \& Burkepile, J. T. 2001, \apj, 549, 1121

%\bibitem[Gilbert et al. (2007)]{Gilbert 2007}
%Gilbert, H. R., Alexander, D., \& Liu, R. 2007, \solphys, 245, 287

%\bibitem[Golub et al. (2007)]{Golub 2007}
%Golub, L., Deluca, E., Austin, G., et al.\ 2007, \solphys, 243, 63

%\bibitem[Golub et al. (2004)]{Golub 2004}
%Golub, L., Deluca, E. E., Sette, A., \& Weber, M. 2004, in ASP
%Conf. Ser. 325, The Solar-B Mission and the Forefront of Solar
%Physics, ed. T. Sakurai \& T. Sekii (San Francisco, CA: ASP), 217

\bibitem[Gopalswamy et al. (2003)]{Goplaswamy 2003}
Gopalswamy, N., Shimojo, M., Lu, W., et al. 2003, \apj, 586, 562

%\bibitem[Gosling et al.(1976)]{1976SoPh...48..389G}
%Gosling, J.~T., Hildner, E., MacQueen, R.~M., et al.\ 1976,
%\solphys, 48, 389

%\bibitem[Gosling \& McComas (1987)]{Gosling 1987}
%Gosling, J. T., \& McComas, D. J., et al.\ 1987, \grl, 14, 335

\bibitem[Gosling et al. (1991)]{Gosling 1991}
Gosling, J. T., McComas, D. J., Phillips, J. L., \& Bame, S. J.
1991, \jgr, 96, 7831

%\bibitem[Green et al. (2007)]{Green 2007}
%Green, L. M., Kliem, B., T\" or\" ok, T., van Driel-Gesztelyi, L.,
%\& Attrill, G. D. R. 2007, \solphys, 246, 365

%\bibitem[Gosling (1993)]{Gosling 1993} Gosling, J. T. 1993, \jgr,
%98, 18937

%\bibitem[Gosling (1991)]{Gosling 1991}
%Gosling, J. T., McComas, D. J., Phillips, J. L., \& Bame, S. J.
%1991, \jgr, 96, 7831

%\bibitem[Gou et al. (2019)]{Gou Tingyu 2019}
%Gou, T. Y., Liu R., Kliem, B., Wang, Y. M., \& Veronig, A. M.
%2019, Sci. Adv. 5, eaau7004

%\bibitem[Harrison(1995)]{1995A&A...304..585H}
%Harrison, R.~A.\ 1995, \aap, 304, 585

%\bibitem[Haw et al. 2018]{Haw 2018 cavity formation}
%Haw, M. A., Wongwaitayakornkul, P., Li, H. \& Bellan, P. M. 2018,
%\apjl, 862, L15

\bibitem[Hayes et al. (2001)]{electron density and brightness}
Hayes, A. P., Vourlidas, A., \& Howard, R. A., 2001, \apj, 548,
1081

%\bibitem[Hirayama(1974)]{1974SoPh...34..323H}
%Hirayama, T.\ 1974, \solphys, 34, 323

%\bibitem[Hoeksema et al. (2014)]{Hoeksema 2014}
%Hoeksema, J. T., Liu, Y., Hayashi, K., et al. 2014, \solphys, 289,
%3483

\bibitem[Howard et al. 2012]{Howard 2012 cavity to be the MFR}
Howard, T. A., \& DeForest, C. E. 2012, \apj, 746, 64

\bibitem[Howard et al. 2017]{Howard 2017 challening the cavity}
Howard, T. A., DeForest, C. E., Schneck, U. G., \& Alden, C. R.
2017, \apj, 834, 86

\bibitem[Howard et al. (2008)]{Howard 2008}
Howard, R. A., Moses, J. D., Vourlidas, A., et al. 2008, \ssr,
136, 67

%\bibitem[Hu et al.(2003)]{2003JGRA..108.1072H}
%Hu, Y.~Q., Li, G.~Q.,\& Xing, X.~Y.\ 2003, Journal of Geophysical
%Research (Space Physics), 108, 1072

\bibitem[Illing \& Hundhausen (1983)]{Illing 1983}
Illing, R. M. E., \& Hundhausen, A. J. 1983, \jgr, 88, 10210

%\bibitem[Illing \& Hundhausen (1985)]{Illing 1985}
%Illing, R. M. E., \& Hundhausen, A. J. 1985, \jgr, 90, 275

%\bibitem[Isavnin et al. (2014)]{Isavnin 2014}
%Isavnin, A., Vourlidas, A., \& Kilpua, E. K. J.\ 2014, \solphys,
%289, 2141

%\bibitem[Isenberg \& Forbes (2007)]{Isenberg 2007}
%Isenberg, P.~A., \& Forbes, T.~G. \ 2007, \apj, 670, 1453

%\bibitem[Isenberg et al. (1993)]{Isenberg 1993}
%Isenberg, P. A., Forbes, T. G., \& Demoulin, P. 1993, \apj, 417,
%368

%\bibitem[Ji et al. (2003)]{Ji Haisheng 2003}
%Ji, H. S., Wang, H. M., Schmahl, E. J., et al. 2003 \apj, 595,
%L135

%\bibitem[Jing et al. (2005)]{Jing 2005}
%Jing, J., Qiu, J., Lin, J., et al.\ 2005, \apj, 620, 1085

%\bibitem[Ji et al. (2003)]{Ji 2003}
%Ji, H. S, Wang, H. M., Schmahl, E. J., Moon, Y. J., \& Jiang, Y. C. 2003, \apj, 595, L135

%\bibitem[Jiang et al. (2009)]{Jiang 2009}
%Jiang, Y. C., Yang, J. Y., Zheng, R. S., Bi, Y., \& Yang, X. L. 2009, \apj, 693, 1851

%\bibitem[Judge (2010)]{Judge 2010}
%Judge, P. G. 2010, \apj, 708, 1238

%\bibitem[Kano et al. (2008)]{Kano 2008}
%Kano, R., Sakao, T., Hara, H., et al.\ 2008, \solphys, 249, 263

%\bibitem[Kazachenko et al. (2017)]{Kazachenko 2017}
%Kazachenko, M. D., Lynch, B. J., Welsch, B. T., \& Sun, X. D.\
%2017, \apj, 845, 49

%\bibitem[Kliem \& T\" or\" ok (2006)]{2006PhRvL..96y5002K}
%Kliem, B., \& T\" or\" ok, T.\ 2006, Physical Review Letters, 96,
%255002

%\bibitem[Kliem et al. (2012)]{Kliem 2012}
%Kliem, B., T\" or\" ok, T., \& Thompson, W. T.\ 2012, \solphys,
%281, 137

%\bibitem[Kopp\& Pneuman(1976)]{1976SoPh...50...85K}
%Kopp, R.~A., \& Pneuman, G.~W.\ 1976, \solphys, 50, 85

%\bibitem[Kosugi et al. (2007)]{Kosugi 2007} Kosugi, T.,
%Matsuzaki, K., Sakao, T., et al.\ 2007, \solphys, 243, 3

%\bibitem[Kahler et al. (1989)]{Kahler 1989}
%Kahler, S. W., Sheeley, N. R., Jr., \& Liggett, M. 1989, \apj, 344, 1026

%\bibitem[Kaiser et al. (2007)]{Kaiser 2007}
%Kaiser, M. L., Kucera, T. A., Davila, J. M., et al. 2007, \ssr, 136, 5

\bibitem[Kippenhahn \& Schl\"uer 1957]{Kippenhahn 1957}
Kippenhahn, R., \& Schl\"uer A. 1957, ZA, 43, 36

%\bibitem[Kliem \& T\" or\" ok (2006)]{Kliem 2006}
%Kliem, B., \& T\" or\" ok, T. 2006, \prl, 96, 255002

%\bibitem[Klimchuk (2001)]{Klimchuk 2001}
%Klimchuk, J. A. 2001, in Space Weather, ed. Song, P., Singer, H.,
%\& Siscore, G. (Geophysical Monograph 125; Washington, DC: Am.
%Geophys. Un.), 143

%\bibitem[Kopp \& Pneuman (1976)]{Kopp 1976}
%Kopp, R. A., \& Pneuman, G. W. 1976, \solphys, 50, 85

%\bibitem[Kuridze et al. (2013)]{Kuridze 2013}
%Kuridze, D., Mathioudakis, M., Kowalski, A. F., et al. 2013, \aap, 552, A55

\bibitem[Lemen et al.(2012)]{2012SoPh..275...17L}
Lemen, J.~R., Title, A.~M., Akin, D.~J., et al.\ 2012, \solphys,
275, 17

%\bibitem[Liu et al. (2018)]{Liu Ying CME rotation 2018 ApJ}
%Liu, Y. A., Liu, Y. D., Hu, H. D., Wang, R., \& Zhao, X. W. 2018,
%\apj, 854, 126

\bibitem[Lin \& Forbes (2000)]{Lin 2000}
Lin, J., \& Forbes, T. G. 2000, \jgr, 105, 2375

\bibitem[Lin et al. (2004)]{Lin 2004}
Lin, J., Raymond, J. C., \& van Ballegooijen, A. A. 2004, \apj£¬
602, 422

%\bibitem[Lin et al. (2003)]{Lin 2003}
%Lin, J., Soon, W., Baliunas, S. L.\ 2003, New Astronomy Reviews,
%47, 53

%\bibitem[Lin et al.(2002)]{2002SoPh..210....3L}
%Lin, R.~P., Dennis, B.~R., Hurford, G.~J., et al.\ 2002, \solphys,
%210, 3

%\bibitem[Lin et al.(2003)]{Lin 2003}
%Lin, R.~P., Soon, W., \& Baliunas, S. L.\ 2003, NewAR, 47, 53

%\bibitem[Linker et al. (2001)]{Linker 2001}
%Linker, J. A., Lionello, R., \& Miki\'{c}, Z. 2001, \jgr, 106,
%25165

%\bibitem[Landi et al. (2010)]{Landi 2010}
%Landi, E., Raymond, J. C., Miralles, M. P., \& Hara, H. 2010 \apj, 711, 75

%\bibitem[Lemen et al. (2012)]{Lemen 2012}
%Lemen, J. R., Title, A. M., Akin, D. J., et al. 2012, \solphys,
%275, 17

%\bibitem[Lepping et al. (1990)]{Lepping 1990}
%Lepping, R. P., Burlaga, L. F., \& Jones, J. A. 1990,\jgr, 95, 11957

%\bibitem[Li \& Zhang (2013)]{Li 2013}
%Li, L. P., \& Zhang, J. 2013, \aap, 552, L11

%\bibitem[Lin et al. (2002)]{Lin 2002}
%Lin, R. P., Dennis, B. R., Hurford, G. J. et al. 2002, \solphys, 210, 3

%\bibitem[Liu et al. (2012)]{Liu 2012}
%Liu, R., Kliem, B., T\" or\" ok, T., et al. 2012, \apj, 756, 59.

%\bibitem[Liu et al. (2008)]{Liu 2008}
%Liu, W., Petrosian, V., Dennis, B. R., \& Jiang, Y. W. 2008, \apj, 676, 704

\bibitem[Liu et al. (2009)]{filament to core}
Liu, Y., Davies, J. A., Luhmann, J. G., et al. 2010, \apjl, 710,
L82

%\bibitem[Liu et al (2009)]{Liu 2009}
%Liu, Y., Su, J., Xu, Z., et al. 2009, \apj, 696, L70

\bibitem[Liu et al. (2014)]{filament to core}
Liu, Y. D., Luhmann, J. G., Kajdic, P., et al. 2014, Nature
Communications, 5, 3481

\bibitem[Liu et al. (2019)]{liu ying}%{c170910 event c}
Liu, Y. D., Zhu, B., \& Zhao, X. W. 2019, \apj, 871, 8

%\bibitem[Lynch et al. 2008]{Lynch 2008}
%Lynch, B. J., Antiochos, S. K., Devore, C. R., Luhmann, J. G., \&
%Zurbuchen, T. H. 2008, \apj, 683, 1192

%\bibitem[Lynch et al. 2004]{Lynch 2004}
%Lynch, B. J., Antiochos, S. K., MacNeice, P. J., Zurbuchen, T. H.,
%\& Fisk, L. A. 2004, \apj, 617, 589


%\bibitem[Lugaz et al. (2011)]{Lugaz 2011}
%Lugaz, N., Downs, C., Shibata, K., et al. 2011, \apj, 738, 127

%\bibitem[Ma et al. (2011)]{Ma 2011}
%Ma, S. L., Raymond, J. C., Golub, L., et al. 2011, \apj, 738, 160

%\bibitem[MacQueen et al. (1986)]{MacQueen 1986}
%MacQueen, R. M., Hundhausen, A. J., \& Conover, C. W.\ 1986, \jgr,
%91, 31


%\bibitem[Manchester et al. (2017)]{Manchester 2017}
%Manchester, I. W., Kilpua, E. K. J., Liu, Y. D. et al.\ 2017,
%\ssr, 212, 1159

%\bibitem[Mari\v ci\'c et al.(2007)]{2007SoPh..241...99M}
%Mari\v ci\'c, D., Vr\v snak, B., Stanger, A.~L., et al.\ 2007,
%\solphys, 241, 99

%\bibitem[Martin (2003)]{Martin 2003}
%Martin, S. F. 2003, Adv. Space Res., 32, 1883

%\bibitem[McKenzie\& Canfield(2008)]{2008A&A...481L..65M}
%McKenzie, D.~E., \&Canfield, R.~C.\ 2008, \aap, 481, L65

\bibitem[Miki\'{c} \& Linker (1994)]{Mikic 1994}
Miki\'{c}, Z., \& Linker, J. A. 1994, \apj, 430, 898

%\bibitem[Miklenic et al. (2009)]{Miklenic 2009}
%Miklenic, C. H., Veronig, A. M., \& Vr\v snak, B., 2009, \aap,
%499, 893

%\bibitem[Miklenic et al. (2007)]{Miklenic 2009}
%Miklenic, C. H., Veronig, A. M., Vr\v snak, B. \& Hanslmeier, A.,
%2007, \aap, 461, 697

%\bibitem[Mostl et al. 2015]{Mostl 2015}
%M\"{o}stl, C., Rollett, T., Frahm, R. A., et al. 2015, Nat.
%Commun. 6, 7135

%\bibitem[Muglach et al. (2009)]{Muglach 2009}
%Muglach, K., Wang, Y. M., \& Kliem, B. 2009, \apj, 703, 976

%\bibitem[Munro, et al. 1979]{Munro 1979}
%Munro, R. H., Gosling, J. T., Hildner, E., et al. 1979, \solphys,
%61, 201

%\bibitem[Neupert (1968)]{Neupert 1968}
%Neupert, W. M., 1968, \apjl, 153, L59

\bibitem[Nindos et al. (2015)]{Nindos 2015}
Nindos, A., Patsourakos, S., Vourlidas, A., \& Tagikas, C. 2015,
\apj, 808, 117

%\bibitem[Olmedo\& Zhang(2010)]{2010ApJ...718..433O}
%Olmedo, O., \& Zhang, J.\ 2010, \apj, 718, 433

%\bibitem[Ouyang et al. (2015)]{Ouyang et al. 2015}
%Ouyang, Y., Yang, K., \& Chen, P. F. 2015, \apj, 815, 72

%\bibitem[Ouyang et al. (2017)]{Ouyang et al. 2017}
%Ouyang, Y., Zhou, Y. H., Chen, P. F., \& Fang, C. 2017, \apj, 835,
%94

%\bibitem[O'Dwyer et al. (2010)]{O'Dwyer 2010}
%O'Dwyer, B., Del Zanna, G., Mason, H. E., Weber, M. A.,
%\&Tripathi, D. 2010, \aap, 521, A21

%\bibitem[Olmedo \& Zhang (2010)]{Olmedo 2010}
%Olmedo, O., \& Zhang, J. 2010, \apj, 718, 433

%\bibitem[Panasenco et al. (2011)]{Panasenco 2011}
%Panasenco, O., Martin, S., Joshi, A. D., \& Srivastava, N. 2011,
%J. Atmos. Solar-Terr. Phys. 73, 1129

%\bibitem[Pneuman (1980)]{Pneuman 1980}
%Pneuman, G. W. \ 1980, \solphys, 65, 369

%\bibitem[Priest\& Forbes(1990)]{1990SoPh..126..319P}
%Priest, E.~R., \& Forbes, T.~G.\ 1990, \solphys, 126, 319

%\bibitem[Priest\& Forbes(2002)]{Priest 2002}
%Priest, E.~R., \& Forbes, T.~G.\ 2002, A\&ARv, 10, 313

\bibitem[Patsourakos et al. (2013)]{Patsourakos 2013}
Patsourakos, S., Vourlidas, A., \& Stenborg, G. 2013, \apj, 764,
125

%\bibitem[Pesnell et al. (2012)]{Pesnell 2012}
%Pesnell, W. Dean, Thompson, B. J., \& Chamberlin, P. C. 2012, \solphys, 275, 3

%\bibitem[Priest \& Forbes (2000)]{Priest 2000}
%Priest, E., \& Forbes, T. 2000, Magnetic Reconnection (Cambridge: Cambridge
%Univ. Press)

%\bibitem[Qiu et al.(2007)]{2007ApJ...659..758Q}
%Qiu, J., Hu, Q., Howard, T.~A., \& Yurchyshyn, V.~B.\ 2007, \apj,
%659, 758

%\bibitem[Qiu et al.(2004)]{2004ApJ...604..900Q}
%Qiu, J., Wang, H., Cheng, C.~Z., \& Gary, D.~E.\ 2004, \apj, 604,
%900

%\bibitem[Qiu \& Yurchyshyn (2005)]{Qiu 2005}
%Qiu, J., \& Yurchyshyn, V. B. 2005, \apj, 634, L121

%\bibitem[Reeves et al. (2010)]{Reeves 2010}
%Reeves, K. K., Linker, J. A., Miki\'{c}, Z., \& Forbes, T. G.
%2010, \apj, 721, 1547

%\bibitem[Reva et al. (2017)]{Reva 2017}
%Reva, A. A., Kirichenko, A. S., Ulyanov, A. S., \& Kuzin, S. V.,
%2017, \apj, 851, 108

%\bibitem[Roussev et al.(2012)]{2012NatPh...8..845R}
%Roussev, I.~I., Galsgaard, K., Downs, C., et al.\ 2012, Nature
%Physics, 8, 845

%\bibitem[Rust (2003)]{Rust 2003}
%Rust, D. M. 2003, Adv. Space Res., 32, 1895

%\bibitem[Rust \& LaBonte (2005)]{Rust 2005}
%Rust, D. M., \& LaBonte, B. J. 2005, \apj, 622, L69

%\bibitem[Sakurai (1976)]{Sakurai 1976}
%Sakurai, T. 1976, PASJ, 28, 177

%\bibitem[Sturrock et al. (2001)]{Sturrock 2001}
%Sturrock, P. A., Weber, M., Wheatland, M. S., \& Wolfson, R. 2001,
%\apj, 548, 492

%\bibitem[Rust\& Kumar(1994)]{1994SoPh..155...69R}
%Rust, D.~M., \& Kumar, A.\ 1994, \solphys, 155, 69

%\bibitem[Schmieder et al. (2015)]{Schmieder 2015}
%Schmieder, B., Aulanier, G., \& Vr\v snak, B.\ 2015, \solphys,
%290, 3457

%\bibitem[Sheeley et al.(1999)]{1999JGR...10424739S}
%Sheeley, N.~R., Walters, J.~H., Wang, Y.-M., \& Howard, R.~A.\
%1999, \jgr, 104, 24739

%\bibitem[Reeves \& Weber (2009)]{Reeves 2009}
%Reeves, K. K., \& Weber, M. A. 2009, in ASP Conf. Ser. 415, The Second
%Hinode Science Meeting: Beyond Discovery-Toward Understanding, ed.
%B. Lites, M. Cheung, T. Magara, J. Mariska, \& K. Reeves (San Francisco,
%CA: ASP), 443

%\bibitem[Rust \& Kumar (1994)]{Rust 1994}
%Rust, D. M., \& Kumar, A. 1994, \solphys, 155, 69

%\bibitem[Rust \& LaBonte (2005)]{Rust 2005}
%Rust, D. M., \& LaBonte, B. J. 2005, \apj, 622, L69

\bibitem[Seaton \& Darnel]{SUVI instrument}
Seaton, D. B., \& Darnel, J. M. 2018, \apjl, 852, L9

%\bibitem[Shen et al. (2011)]{Shen ChengLong 2011}
%Shen, C. L., Wang, Y. M., Gui, B., Ye, P. Z., \& Wang, S. 2011,
%\solphys, 269, 389

%\bibitem[Song et al.(2017a)]{Song 2017a origin of filament}
%Song, H.~Q., Chen, Y., Li, B., et al.\ 2017a, \apjl, 836, L11

%\bibitem[Song et al.(2018)]{2018ApJ...857..L21}
%Song, H.~Q., Chen, Y., Qiu, J., et al.\ 2018, \apjl, 857, L21

%\bibitem[Song et al.(2013)]{2013ApJ...773..129S}
%Song, H.~Q., Chen, Y., Ye, D.~D., et al.\ 2013, \apj, 773, 129

%\bibitem[Song et al.(2015a)]{Song 2015}
%Song, H.~Q., Chen, Y., Zhang, J., et al.\ 2015, \apj, 804, L38

\bibitem[Song et al.(2015b)]{Song 2015b, hc as the MFR}
Song, H.~Q., Chen, Y., Zhang, J., et al.\ 2015, \apjl, 808, L15

\bibitem[Song et al. (2017b)]{Song 2017 cme three-part}
Song, H. Q., Cheng, X., Chen, Y. et al., 2017, \apj, 848, 21

\bibitem[Song et al.(2014a)]{2014ApJ...792L..40S Flux rope formation}
Song, H.~Q., Zhang, J., Chen, Y., \& Cheng, X.\ 2014a, \apjl, 792,
L40

\bibitem[Song et al.(2014b)]{2014ApJ...784...48S temperature evolution}
Song, H.~Q., Zhang, J., Cheng, X., et al.\ 2014b, \apj, 784, 48

\bibitem[Song et al. (2019)]{double cores}
Song, H. Q., Zhang, J., Cheng, X., et al. 2019, \apj, 883, 43

%\bibitem[Sturrock(1968)]{1968IAUS...35..471S}
%Sturrock, P.~A.\ 1968, Structure and Development of Solar Active
%Regions, 35, 471

%\bibitem[Su et al.(2012)]{2012ApJ...746L...5S}
%Su, Y., Dennis, B.~R., Holman, G.~D., et al.\ 2012, \apjl, 746, L5

%\bibitem[Schmelz et al. (2011a)]{Schmelz 2011a}
%Schmelz, J. T., Rightmire, L. A., Saar, S. H., et al. 2011a, \apj,
%738, 146

%\bibitem[Schmelz et al. (2010)]{Schmelz 2010}
%Schmelz, J. T., Saar, S. H., Nasraoui, K. et al. 2010, \apj, 723,
%1180

%\bibitem[Schmelz et al. (2011b)]{Schmelz 2011b}
%Schmelz, J. T., Worley, B. T., Anderson, D. J., et al. 2011b,
%\apj, 739, 33

%\bibitem[Song et al. (2013)]{Song 2013}
%Song, H. Q., Chen, Y., Ye, D. D., et al. 2013, \apj, 773, 129


%\bibitem[Song et al. (2009)]{Song 2009}
%Song, H. Q., Chen, Y., Liu, K., Feng, S. W., \& Xia, L. D. 2009,
%\solphys, 258, 129

%\bibitem[Song et al. (2012)]{Song 2012}
%Song, H. Q., Kong, X. L., Chen, Y., et al. 2012, \solphys, 2012,
%276, 261

%\bibitem[Song et al. (2014)]{Song 2014}
%Song, H. Q., Zhang, J., Cheng, X., et al. 2014, \apj, 784, 48

%\bibitem[Sturrock (1966)]{Sturrock 1966}
%Sturrock, P. A. 1966, Nature, 211, 695

%\bibitem[Su et al. (2011)]{Su 2011}
%Su, Y., Surges, V., van Ballegooijen, A., DeLuca, E., \& Golub, L.
%2011, \apj, 734, 53

%\bibitem[Summers (1974)]{Summers 1974}
%Summers, H. P. 1974, \mnras, 169, 663

%\bibitem[Temmer et al. (2010)]{Temmer 2010}
%Temmer, M., Veronig, A. M., Kontar, E. P., Krucker, S., \& Vr\v
%snak, B. 2010, \apj, 712, 1410

%\bibitem[Temmer et al. (2008)]{Temmer 2008}
%Temmer, M., Veronig, A. M., Vr\v snak, B. et al. 2008, \apj, 673,
%L95

%\bibitem[Thompson (2011)]{Thompson 2011}
%Thompson, W. T. 2011, Journal of Atmospheric and Solar-Terrestrial
%Physics, 73, 1138

%\bibitem[Thompson et al. (2012)]{Thompson et al. 2012}
%Thompson, W. T., Kliem, B., T\" or\" ok, T. 2012, \solphys, 276,
%241

%\bibitem[Tian et al (2012)]{Tian 2012}
%Tian, H., McIntosh, S. W., Xia, L. D., He, J. S., \& Wang, X. 2012 \apj, 748, 106

%\bibitem[Titov\& D{\'e}moulin(1999)]{1999A&A...351..707T}
%Titov, V.~S., \&D{\'e}moulin, P.\ 1999, \aap, 351, 707

%\bibitem[Torok et al. (2010)]{Torok 2010}
%T\" or\" ok, T., Berger, M. A., \& Kliem, B.\ 2010, \aap, 516, A49

%\bibitem[Torok \& Kliem. (2005)]{Torok 2004}
%T\" or\" ok, T., \& Kliem, B. 2005, \apjl, 630, L97

%\bibitem[Torok et al. (2004)]{Torok 2004}
%T\" or\" ok, T., Kliem, B. \& Titov, V. S. 2004, \aap, 413, L27

%\bibitem[van Tend\& Kuperus(1978)]{1978SoPh...59..115V}
%van Tend, W., \& Kuperus, M.\ 1978, \solphys, 59, 115

\bibitem[Veronig et al. (2018)]{20170910 event}
Veronig, A. M., Podladchikova, T., Dissauer, K., et al. 2018,
\apj, 868, 107

%\bibitem[Vr\v snak  (1990)]{Vrsnak 1990}
%Vr\v snak, B. 1990, \solphys, 129, 295

%\bibitem[Vr\v snak  (2008)]{Vrsnak 2008}
%Vr\v snak, B. 2008, Ann. Geophys., 26, 3089

%\bibitem[Vr\v snak  (2016)]{Vrsnak 2016}
%Vr\v snak, B. 2016, Astron. Nachr., 337, 1002

%\bibitem[Vr\v snak et al. (2005)]{Vrsnak 2005}
%Vr\v snak, B., Sudar, D., \& Ru\v zdjak, D. 2005, \aa, 435, 1149

%\bibitem[V\'{a}squez et al (2010)]{Vasquez 2010}
%V\'{a}squez, A. M., Frazin, R. A., \& Manchester, W. B., IV. 2010, \apj, 715, 1352

\bibitem[Vourlidas (2006)]{Vourlidas 2006 white-light brightness correlated with the POS}
Vourlidas, A., \& Howard, R. A. 2006, \apj, 642, 1216

\bibitem[Vourlidas et al. (2013)]{Vourlidas 2013}
Vourlidas, A., Lynch, B. J., Howard, R. A., \& Li, Y. 2013,
\solphys, 284, 179

%\bibitem[Wang et al. (2003)]{Wang 2003}
%Wang, H. M., Qiu, J., Jing, J., \& Zhang H. Q. 2003, \apj, 593,
%564

%\bibitem[Wang et al. (2017)]{Wang wensi 2017}
%Wang, W. S., Liu, R., Wang, Y. M., et al. 2017, Nature
%Communications, 8, 1330

%\bibitem[Wang et al. (2004)]{Wang YiMin 2004}
%Wang, Y. M., Shen, C. L., Wang, S., \& Ye, P. 2004, \solphys, 222,
%329

%\bibitem[Wang\& Stenborg(2010)]{2010ApJ...719L.181W}
%Wang, Y.-M., \& Stenborg, G.\ 2010, \apjl, 719, L181


%\bibitem[Wang\& Zhang(2007)]{2007ApJ...665.1428W}
%Wang, Y., \& Zhang, J.\ 2007, \apj, 665, 1428

\bibitem[Webb et al. (1994)]{Webb 1994}
Webb, D. F., Forbes, T. G., Aurass, H., et al. 1994, \solphys,
153, 73

\bibitem[Webb \& Howard (2012)]{Webb and Howard 2012 Living Reviews in Solar
Physics}
Webb, D. F., \& Howard, T. A. 2012, LRSP, 9, 3

%\bibitem[Webb \& Hundhausen (1987)]{Webb 1987}
%Webb, D.F., \& Hundhausen, A.J. 1987, \solphys, 108, 383

\bibitem[Webb et al. (2000)]{Webb 2000 space weather effect}
Webb, D. F., Lepping, R. P., Burlaga, L. F., et al. 2000, \jgr,
105, 27251

%\bibitem[Williams et al. (2005)]{Williams 2005}
%Williams, D. R., T\" or\" ok, T., D\'{e}moulin, P., van
%Driel-Gesztelyi, L., \& Kliem, B. 2005, \apjl, 628, L163

%\bibitem[Wang \& Stenborg, G. (2010)]{Wang 2010}
%Wang, Y. M., \& Stenborg, G. 2010, \apj, 719, L181

%\bibitem[Warren et al. (2013)]{Warren 2013}
%Warren, H. P., Mariska, J. T., \& Doschek, G. A. 2013, \apj, 770, 116

%\bibitem[Wang \& Zhang (2007)]{Wang 2007}
%Wang, Y. M., \& Zhang, J. 2007, \apj, 665, 1428


%\bibitem[Weber et al. (2004)]{Weber 2004}
%Weber, M. A., Deluca, E. E., Golub, L., \& Sette, A. L. 2004, in
%IAU Symp. 223, Multi-Wavelength Investigations of Solar Activity,
%ed. A. V. Stepanov, E. E. Benevolenskaya, \& A. G. Kosovichev
%(Cambridge: Cambridge Univ. Press), 223, 321

%\bibitem[Williams et al. (2005)]{Williams 2005}
%Williams, D. R., T\" or\" ok, T., D\'{e}moulin, P., van Driel-Gesztelyi, L., \& Kliem, B. 2005, \apj, 628, L163

%\bibitem[Winebarger et al. (2011)]{Winebarger 2011}
%Winebarger, A., Schmelz, J., Warren, H., Saar, S., \& Kashyap, V.
%2011, \apj, 740,2

%\bibitem[Wood et al. (2016)]{filament concentrate}
%Wood, B. E., Howard, R. A., \& Linton, M. G. 2016, \apj, 816, 67

%\bibitem[Yan et al. (2016)]{Filament formation}
%Yan, X. L., Priest, E. R., Guo, Q. L., et al. 2016, \apj, 832, 23

%\bibitem[Yan et al. (2014)]{Yan et al. Filament rotation and Kink Instability 2014}
%Yan, X. L., Xue, Z. K., Liu, J. H., et al. 2014, \apj, 782, 67

\bibitem[Yan et al. (2018)]{yan xiao li}%{b 20170910event b}
Yan, X. L., Yang, L. H., Xue, Z. K., et al. 2018, \apj, 883, L18

%\bibitem[Yashiro et al.(2005)]{2005JGRA..11012S05Y}
%Yashiro, S., Gopalswamy, N., Akiyama, S., Michalek, G., \& Howard,
%R.~A.\ 2005, Journal of Geophysical Research (Space Physics), 110,
%A12S05

%\bibitem[Yashiro et al. (2002)]{Yashiro 2002}
%Yashiro, S., Gopalswamy, N., Michalek, G., \& Howard, R. A. 2002,
%BAAS, 34, 695

%\bibitem[Yashiro et al. (2004)]{Yashiro 2004}
%Yashiro, S., Gopalswamy, N., Michalek, G., et al. 2004, \jgr, 109,
%7105

%\bibitem[Yurchyshyn (2008)]{Yurchyshyn 2008}
%Yurchyshyn, V. 2008, \apjl, 675, L49

\bibitem[Zhang et al.(2012)]{2012NatCo...3E.747Z}
Zhang, J., Cheng, X., \& Ding, M.-D.\ 2012, Nature Communications,
3, 747

%\bibitem[Zhang \& Dere. (2006)]{Zhang 2006}
%Zhang, J., \& Dere, K. P. 2006, \apj, 649, 1100

\bibitem[Zhang et al.(2003)]{2003ApJ...582..520Z}
Zhang, J., Dere, K.~P., Howard, R.~A., \& Bothmer, V.\ 2003, \apj,
582, 520

%\bibitem[Zhang et al.(2001)]{2001ApJ...559..452Z}
%Zhang, J., Dere, K.~P., Howard, R.~A., Kundu, M.~R., \& White,
%S.~M.\ 2001, \apj, 559, 452

%\bibitem[Zhang et al. (2004)]{Zhang 2004}
%Zhang, J., Dere, K. P., Howard, R. A., \& Vourlidas, A. 2004, \apj, 604, 420

%\bibitem[Zhang et al. (2013)]{Zhang 2013}
%Zhang, J., Hess, P., \& Poomvises, W. 2013, \solphys, 284, 89

\bibitem[Zhang et al.(2007)]{2007JGRA..11210102Z}
Zhang, J., Richardson, I.~G., Webb, D.~F., et al.\ 2007, \jgr,
112, A10102

\bibitem[Zhang Jun (2015)]{Flux rope statistics}
Zhang, J., Yang, S. H., \& Li, T. 2015, \aap, 580, A2

%\bibitem[Zhou et al. (2006)]{Zhou et al. 2006}
%Zhou, G. P., Wang, J. X., Zhang, J., et al. 2006, \apj, 651, 1238

%\bibitem[Zhukov \& Auch\`{e}re]{Zhukov 2004}
%Zhukov, A. N., \& Auch\`{e}re, F. 2004 \aap, 427, 705

\bibitem[Zuccarello et al. (2012)]{Zuccarello 2012}
Zuccarello, F. P., Bemporad, A., Jacobs, C., et al.\ 2012, \apj,
744, 66

%========================================================

\end{thebibliography}
\end{document}